\documentclass[twocolumn,tighten]{aastex631}
\usepackage{color}
\usepackage{multirow}

\shortauthors{Sano et al. (2022)}
\shorttitle{An expanding shell of neutral hydrogen associated with SN~1006}

\begin{document}
\title{An Expanding Shell of Neutral Hydrogen Associated with SN~1006: Hints for the Single-Degenerate Origin and Faint Hadronic Gamma-Rays}

\author[0000-0003-2062-5692]{H. Sano}
\affiliation{{Faculty of Engineering, Gifu University, 1-1 Yanagido, Gifu 501-1193, Japan: hsano@gifu-u.ac.jp}}
\affiliation{National Astronomical Observatory of Japan, Mitaka, Tokyo 181-8588, Japan}

\author[0000-0002-5092-6085]{H. Yamaguchi}
\affiliation{Institute of Space and Astronautical Science (ISAS), Japan Aerospace Exploration Agency (JAXA), 3-1-1 Yoshinodai, Chuo-ku, Sagamihara, Kanagawa 252-5210, Japan}
\affiliation{Department of Physics, Graduate School of Science, The University of Tokyo, 7-3-1 Hongo, Bunkyo-ku, Tokyo 113-0033, Japan}

\author[0000-0001-5069-5988]{M. Aruga}
\affiliation{Department of Physics, Nagoya University, Furo-cho, Chikusa-ku, Nagoya 464-8601, Japan}

\author[0000-0002-8966-9856]{Y. Fukui}
\affiliation{Department of Physics, Nagoya University, Furo-cho, Chikusa-ku, Nagoya 464-8601, Japan}

\author[0000-0002-1411-5410]{K. Tachihara}
\affiliation{Department of Physics, Nagoya University, Furo-cho, Chikusa-ku, Nagoya 464-8601, Japan}

\author[0000-0002-4990-9288]{M. D. Filipovi{\'c}}
\affiliation{School of Science, Western Sydney University, Locked Bag 1797, Penrith South DC, NSW 2751, Australia}

\author[0000-0002-9516-1581]{G. Rowell}
\affiliation{School of Physical Sciences, The University of Adelaide, North Terrace, Adelaide, SA 5005, Australia}

\begin{abstract}
We report new H{\sc i} observations of the Type Ia supernova remnant SN~1006 using the Australia Telescope Compact Array with an angular resolution of $4\farcm5 \times 1\farcm4$ ($\sim$2 pc at the assumed SNR distance of 2.2~kpc). We find an expanding gas motion in position--velocity diagrams of H{\sc i} with an expansion velocity of $\sim$4~km~s$^{-1}$ and a mass of $\sim$1000 $M_\odot$. The spatial extent of the expanding shell is roughly the same as that of SN~1006. We here propose a hypothesis that SN~1006 exploded inside the wind-blown bubble formed by accretion winds from the progenitor system consisting of a white dwarf and a companion star, and then the forward shock has already reached the wind wall. This scenario is consistent with the single-degenerate model. We also derived the total energy of cosmic-ray protons $W_\mathrm{p}$ to be only $\sim$1.2--$2.0 \times 10^{47}$ erg by adopting the averaged interstellar proton density of $\sim$25~cm$^{-3}$. The small value is compatible with the relation between the age and $W_\mathrm{p}$ of other gamma-ray supernova remnants with ages below $\sim$6 kyr. The $W_\mathrm{p}$ value in SN~1006 will possibly increase up to several 10$^{49}$ erg in the next $\sim$5 kyr via the cosmic-ray diffusion into the H{\sc i} wind-shell.
\end{abstract}

\section{Introduction}\label{introduction}
Identifying the progenitor system of Type Ia supernovae is one of the important issues of modern astrophysics because of their use as standard candles for measuring the expansion history of the universe \citep[e.g.,][]{1999ApJ...517..565P}. The single-degenerate (SD) and double-degenerate (DD) models are widely accepted to describe the progenitor systems of Type Ia SNe: the SD model in which a white dwarf accreted gaseous materials from a nondegenerate companion until the white dwarf gets close to the Chandrasekhar mass $\sim$1.4 $M_\odot$ \citep{1973ApJ...186.1007W,1982ApJ...257..780N,1984ApJS...54..335I,1985ASSL..113....1P}, and the DD model represents the merger of two white dwarfs \citep{1982ApJ...257..780N,1984ApJ...277..355W}. To distinguish two scenarios, a search for a surviving companion is thought to be essential because it can be seen only in the SD scenario. {Despite many efforts to detect such surviving companions of Type Ia SNRs,} no apparent observational evidence was reported\footnote{{Although a strong candidate for a surviving companion was reported in Tycho's SNR named ``Tycho G'' \citep{2009ApJ...691....1G,2014MNRAS.439..354B,2015ApJ...809..183X,2018MNRAS.479.5696K,2019ApJ...870..135R}, the progenitor system for Tycho's SNR is still being debated due to several significant objections \citep[e.g.,][see also a complete review by \citeauthor{2019NewAR..8501523R} \citeyear{2019NewAR..8501523R}]{2013ApJ...774...99K,2017NatAs...1..800W}.}}{\citep[see reviews by][]{2014ARA&A..52..107M,2016IJMPD..2530024M,2019NewAR..8501523R}}.

An expanding shell (also known as ``wind-blown bubble'') of interstellar neutral gas associated with Type Ia supernova remnants (SNRs) has received much attention as alternative evidence for the SD scenario. Because the expanding gaseous shell could be formed by accretion winds (also known as ``disk wind'' or ``optically-thick wind'') from the progenitor system consisting of a white dwarf and a nondegenerate companion \citep[e.g.,][]{1996ApJ...470L..97H,1999ApJ...522..487H,1999ApJ...519..314H,2008ApJ...679.1390H,2003ApJ...590..445H,2003ApJ...598..527H}, whereas such wind shell is not expected in the DD scenario. The first discovery of such expanding gaseous shell was made by CO observations toward Tycho's SNR \citep{2016ApJ...826...34Z}. The authors argued that the expanding shell with the mass of $\sim$220 $M_\odot$ and an expansion velocity of $\sim$5~km~s$^{-1}$ could be explained by the energy injection from accretion winds, and hence concluded that Tycho's SNR is consistent with the SD scenario. The presence of dense-gas wall and the SD scenario were also supported by the rapid shock deceleration during the last $\sim$15 yr \citep{2021ApJ...906L...3T}. Subsequent CO and H{\sc i} studies found similar expanding shells of atomic and/or molecular clouds in the Type Ia SNRs RCW~86 \citep{2017JHEAp..15....1S}, N103B \citep{2018ApJ...867....7S,2019Ap&SS.364..204A}, and G344.7$-$0.1 \citep{2020ApJ...897...62F}. To better understand the progenitor system of Type Ia supernovae, we need further observations of interstellar molecular and atomic clouds toward other Type Ia SNRs.

SN~1006 (also known as G327.6$+$14.6) is a historical SNR that exploded in AD~1006 \citep{2002ISAA....5.....S}. The small distance of 2.2~kpc from us \citep{2003ApJ...585..324W} is consistent with its young age of $\sim$1000 yr and a large diameter of 28\farcm8 arcmin or $\sim$18 pc. Based on the historical record, SN~1006 is widely thought to originate from a Type Ia supernova \citep{1996ApJ...459..438S}. Owing to its location far from the Galactic plane ($\sim$550 pc), SN~1006 is an ideal object to search for a surviving companion with very little contamination along the line of sight. However, neither non-degenerated companion nor surviving white dwarf companion has been detected to date \citep[e.g.,][]{1980ApJ...241.1039S,2012Natur.489..533G,2012ApJ...759....7K,2018MNRAS.479..192K}. 
{Therefore, SN~1006 is thought to be a remnant that exploded as the DD progenitor system.}

SN~1006 is also noted as an ideal site for cosmic-ray acceleration since the first detection of synchrotron X-rays from the northeast and southwest shells \citep{1995Natur.378..255K}. Subsequent observations of hard X-rays and GeV/TeV gamma-rays suggest the presence of high-energy cosmic-ray electrons up to $\sim$100 TeV \citep[e.g.,][]{2008PASJ...60S.153B,2010A&A...516A..62A,2016ApJ...823...44X,2017ApJ...851..100C,2018ApJ...864...85L}. The latest broadband spectral modeling by \cite{2019PASJ...71...77X} suggests that gamma-ray emission is predominantly the leptonic origin that cosmic-ray electron energies a low-energy photon into the gamma-ray energy via inverse Compton scattering.

The interstellar environments of SN~1006, including both the ionized and neutral gaseous medium, have been well studied by multiwavelength observations covering radio to X-rays. Assuming the standard compression ratio for a strong shock of 4, the optical, infrared, and X-ray observations estimated the pre-shock density of $\sim$0.02--0.4~cm$^{-3}$ from the post-shock electron density \citep[e.g.,][]{1987ApJ...315L.135K,2003ApJ...589..827B,2007A&A...475..883A,2007ApJ...659.1257R,2008PASJ...60S.141Y,2009ApJ...692L.105K,2012A&A...546A..66M,2013ApJ...771...56U,2013ApJ...764..156W,2014ApJ...781...65W,2015MNRAS.453.3953L}. For the neutral hydrogen gas surrounding SN~1006, \cite{2002A&A...387.1047D} carried out H{\sc i} observations with an angular resolution of $4\farcm7 \times 3\farcm0$ (or 3 pc $\times$ 2 pc at the distance of 2.2~kpc). The authors concluded that the H{\sc i} clouds at $V_\mathrm{LSR}$: $-25$ to $-15$~km~s$^{-1}$ are likely interacting with the SNR, and the derived ambient density is $\sim$0.3~cm$^{-3}$. {On the other hand}, \cite{2014ApJ...782L..33M} argued that the H{\sc i} clouds at $V_\mathrm{LSR}$: $\sim$6 to 11~km~s$^{-1}$ are interacting with the southwest shell of the SNR, {by re-analyzing the same H{\sc i} datasets. They also found that the X-ray shell is slightly deformed in the direction of the southwestern H{\sc i} cloud. The spatially-resolved X-ray spectroscopy along the southwestern shell indicated that the X-ray-derived absorbing column density is proportional to the H{\sc i} column densities. Moreover, the cutoff energy of the synchrotron emission decreases in the regions corresponding to the southwestern cloud, suggesting that shock--cloud interaction occurred. Therefore, SN~1006 is a suitable site to test the physical relation among the supernova shocks, ambient clouds, and high-energy radiation.}

Here, we report the spatial and kinematic distributions of H{\sc i} clouds toward SN~1006 using new H{\sc i} observations. Our finding of an expanding H{\sc i} shell provides a unique solution for the cloud association with SN~1006, as well as its progenitor system and cosmic-ray acceleration. In Section~\ref{observations} we present the observations and data reductions. Section~\ref{results} comprises of four subsections: Section~\ref{results:overview} gives an overview of X-rays and H{\sc i} toward SN~1006, Sections \ref{results:channel} and \ref{results:pv} show the spatial and kinematical distributions of H{\sc i} while Section~\ref{results:mass} represents the mass and density of H{\sc i}. In Sections \ref{discussion} and \ref{conclusions} we discuss and conclude our findings.

\section{Observations and Data Reduction}\label{observations}
\subsection{H{\sc i}}\label{observations:hi}
We performed H{\sc i} observations at 1.4 GHz using ATCA, which consists of six 22-m antennas located at Narrabri, Australia. Observations were conducted during 24 hours on November 28, 2013, and March 12, 2014, with ATCA in the EW352 and EW367 array configurations (Project ID: C2857). We employed the mosaicking technique, with seven pointings arranged in a hexagonal grid at the Nyquist spatial separation of $19'$. The absolute flux density was scaled by observing the quasar PKS 0823$-$500, which was used as the primary amplitude and bandpass calibrators. We also periodically observed the quasar PKS 1421$-$490 for gain and phase calibration. We utilized the MIRIAD software package \citep{1995ASPC...77..433S} for the data reduction. To recover extended emission, we combined the ATCA data cube with archival single-dish datasets obtained using the Parkes 64-m radio telescope \citep{2009ApJS..181..398M,2010A&A...521A..17K}. The resulting beam size of H{\sc i} is $4\farcm5 \times 1\farcm4$ with a position angle of $11\fdg5$, corresponding to the spatial resolution of 2.9 pc $\times$ 0.9 pc at an SNR distance of 2.2~kpc. The typical noise fluctuations are 0.32 K per channel for a velocity resolution of 1~km~s$^{-1}$.

\subsection{X-rays}\label{observations:xray}
We used archival X-ray data obtained by Chandra with the Advanced CCD Imaging Spectrometer I-array (Obs IDs: 3838, 4385--4394, 13738--13743, 14423, 14424, and 14435), which have been published by \cite{2008ApJ...680.1180C} and \cite{2014ApJ...781...65W}. We used CIAO version 4.12 \citep{2006SPIE.6270E..1VF} with CALDB 4.9.1 \citep{2007ChNew..14...33G} for data reduction and imaging. After reprocessing for all data using the chandra\_repro task, we created exposure-corrected, energy-filtered maps using the merge\_obs task in the energy bands of 0.5--7.0~keV (broad band), 0.5--1.2~keV (soft band), 1.2--2.0~keV (medium band), and 2.0--7.0~keV (hard band). The resulting effective exposure time is $\sim$800 ks.

\begin{figure*}[]
\begin{center}
\includegraphics[width=\linewidth,clip]{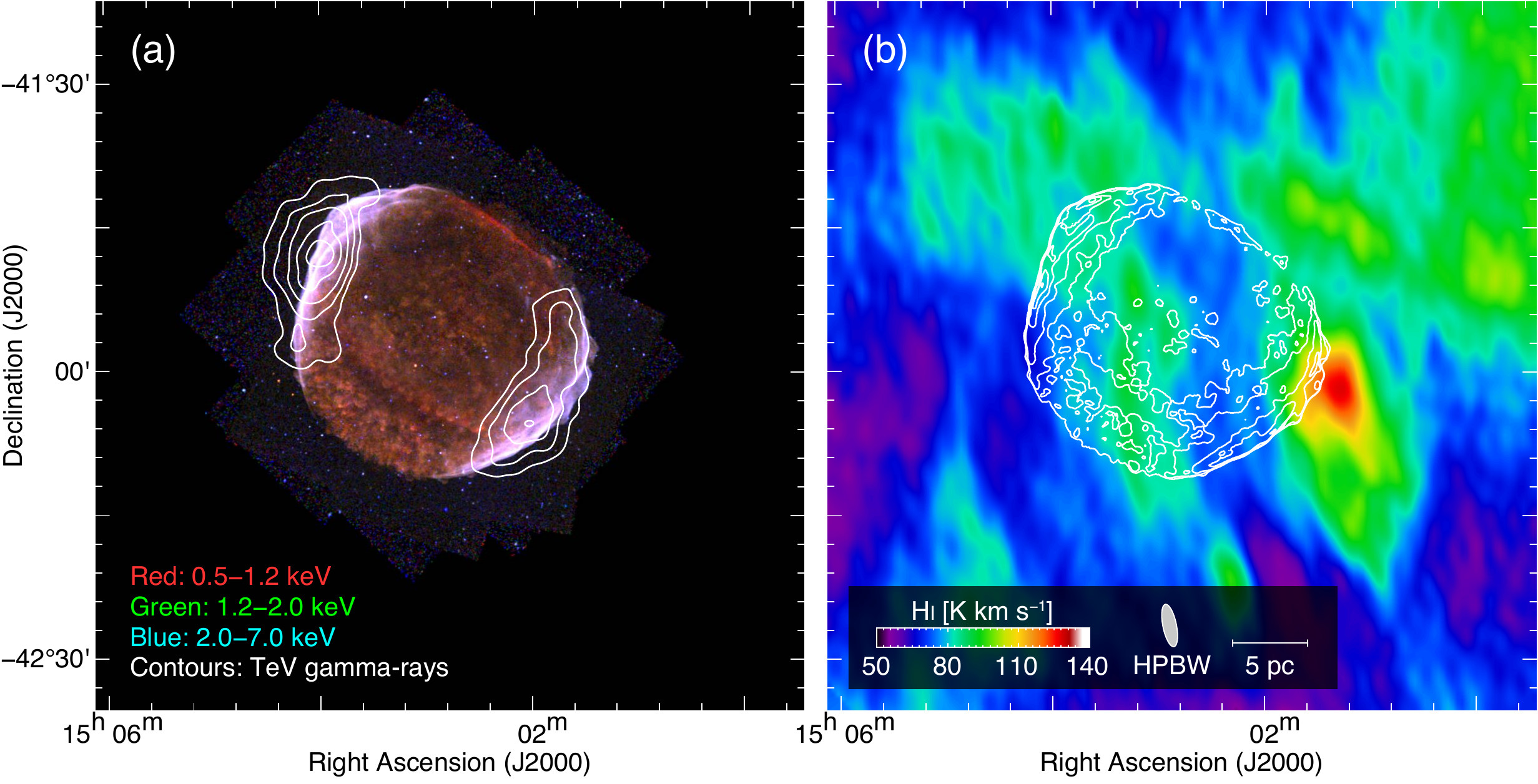}
\caption{(a) RGB X-ray image of SN~1006 obtained with Chandra \citep[][]{2008ApJ...680.1180C,2014ApJ...781...65W}. The red, green, and blue colors correspond to the energy bands 0.5--1.2~keV, 1.2--2.0~keV, and 2.0--7.0~keV, respectively. The superposed contours indicate TeV gamma-ray significance obtained with H.E.S.S. \citep{2010A&A...516A..62A}. The contour levels are 3, 4, 5, 6, and 7 $\sigma$ levels. (b) Velocity integrated intensity map of H{\sc i} obtained with ATCA \& Parkes. The integration velocity range is from 4.0 to 12.0~km~s$^{-1}$. The superposed contours indicate the median-filtered Chandra X-ray intensity in the energy band of 0.5--7.0~keV. The contour levels are 2.5, 4.2, 9.3, 17.8, 29.7, and $45.0 \times 10^{-7}$ photons s$^{-1}$ pixel$^{-1}$.}
\label{fig1}
\end{center}
\end{figure*}%

\section{Results}\label{results}
\subsection{Overview of X-ray and H{\sc i} Distributions}\label{results:overview}
Figure \ref{fig1}(a) shows the false-color image of SN~1006 obtained with Chandra. The X-ray morphology of SN~1006 is that of a nearly circular shell in the soft-band (red: 0.5--1.2~keV), while the medium-band (green: 1.2--2.0~keV) and hard-band (blue: 2.0--7.0~keV) images show strong bilateral symmetry in the northeast and southwest direction. The soft-band image is dominated by thermal X-rays except for the northeast and southwest shells. The brightest northwestern limb is thought to be formed by interactions between the neutral hydrogen gas and supernova shocks \citep[e.g.,][]{2003ApJ...586.1162L,2014ApJ...781...65W}. The hard-band image in the northeast and southwest shells corresponds to non-thermal synchrotron X-rays from cosmic-ray electrons \citep[e.g.,][]{1995Natur.378..255K}, which is also bright in TeV gamma-rays as shown in contours \citep{2010A&A...516A..62A}.

Figure \ref{fig1}(b) shows the integrated intensity map of H{\sc i}. In the present paper, we focus on the velocity range from 4.0 to 12.0~km~s$^{-1}$, which includes the shock-interacting H{\sc i} clouds suggested by \cite{2014ApJ...782L..33M}. We find H{\sc i} clouds not only in the west shell, but also toward the north shell and the center of the SNR. Interestingly, no dense H{\sc i} clouds are adjacent to the {southeast} shells, where the shock velocity shows the maximum value in SN~1006 \citep{2014ApJ...781...65W}. We also note that the H{\sc i} intensity of SN~1006 is about 3--10 times weaker than that of the typical Type Ia SNRs interacting with H{\sc i} clouds in the Galactic plane \citep[e.g.,][]{2017JHEAp..15....1S,2020ApJ...897...62F}.

\subsection{Velocity Channel Distributions of H{\sc i}}\label{results:channel}
\begin{figure*}[]
\begin{center}
\includegraphics[width=\linewidth,clip]{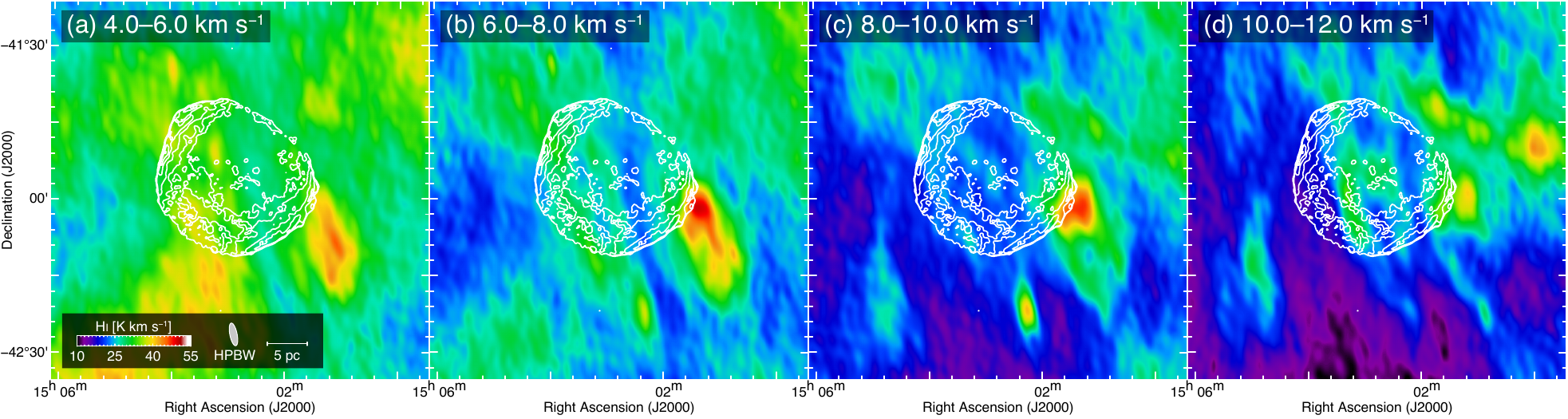}
\caption{Velocity channel distributions of H{\sc i} superposed on the Chandra X-ray contours as shown in Figure \ref{fig1}(b). Each panel shows H{\sc i} intensity map integrated every 2.0~km~s$^{-1}$ in a velocity range from 4.0 to 12.0~km~s$^{-1}$.}
\label{fig2}
\end{center}
\end{figure*}%
Figure \ref{fig2} shows the velocity channel maps of H{\sc i} toward SN~1006. We find diffuse or clumpy H{\sc i} clouds, some of which are along with the X-ray shell boundary. The H{\sc i} clouds at $V_\mathrm{LSR} = 6.0$--8.0~km~s$^{-1}$ lie on the edges of the northeast and southwest X-ray limbs. The northwest shell appears to be associated with H{\sc i} clumps at $V_\mathrm{LSR} = 10.0$--12.0~km~s$^{-1}$. The H{\sc i} intensity at $V_\mathrm{LSR} = 8.0$--10.0~km~s$^{-1}$ decreases toward the center of the SNR, whereas H{\sc i} clouds fill the remnant in the other velocity maps. 

\subsection{Spatial and Kinematic Distributions of H{\sc i}}\label{results:pv}
\begin{figure*}[]
\begin{center}
\includegraphics[width=170mm,clip]{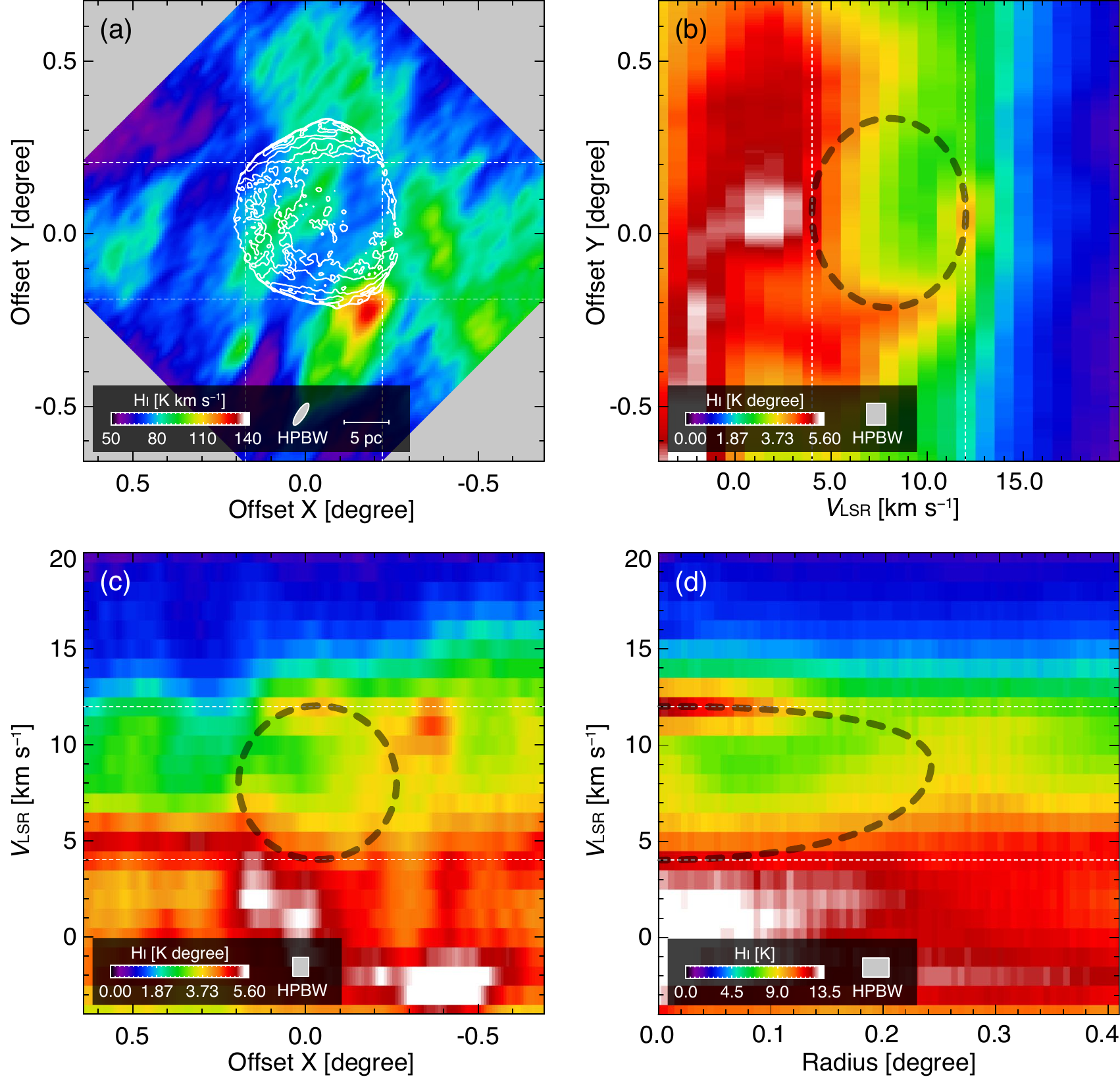}
\caption{(a) Same image and contours of Figure \ref{fig1}(b), but the map was rotated by 45 degrees clockwise. (b--c) Position--velocity (p--v) diagrams of H{\sc i}. The H{\sc i} brightness temperature is averaged from $-$0\fdg22 to 0\fdg17 in Offset-X for (b) and from $-$0\fdg19 to 0\fdg21 in Offset-Y for (c). (d) Radius--velocity (r--v) diagram around the center of the SNR at ($\alpha_\mathrm{J2000}$, $\delta_\mathrm{J2000}) = (15^\mathrm{h}02^\mathrm{m}51\fs1$, $-41\arcdeg55\arcmin32\fs12$). The black dashed circles in the p--v and r--v diagrams indicate the boundaries of the H{\sc i} cavities (see the text). }
\label{fig3}
\end{center}
\end{figure*}
Figures \ref{fig3}(b) and \ref{fig3}(d) show the position--velocity (p--v) diagrams in the Offset X and Y coordinates, which were rotated by 45 degrees clockwise from the equatorial coordinate as shown in Figure \ref{fig3}(a). {Because the H{\sc i} clouds are distributed across the SNR from northeast to southwest, the rotated image is suitable for extracting the p--v diagram along the H{\sc i} distribution.} We find a cavity-like distribution in each p--v diagram of H{\sc i}, whose velocity range is from 4.0 to 12.0~km~s$^{-1}$. {This trend is not significantly changed by varying the integration spatial ranges of Offset X and Y.} It is noteworthy that the spatial extent of each H{\sc i} cavity in the Offset X or Y direction is roughly consistent with the apparent diameter of the X-ray shell. We also calculated an average brightness temperature of H{\sc i} on annuli about the center of the SNR using the tool KSHELL in the KARMA \citep{1996ASPC..101...80G}. Figure \ref{fig3}(d) shows the radius--velocity (r--v) diagram centered at ($\alpha_\mathrm{J2000}$, $\delta_\mathrm{J2000}) = (15^\mathrm{h}02^\mathrm{m}51\fs1$, $-45\arcdeg55'32\farcs12$)\footnote{We used the center position of SN~1006 which was derived by \cite{2010A&A...516A..62A}.}. We find a similar cavity-like distribution of H{\sc i} with the velocity range of $V_\mathrm{LSR}$: 4.0--12.0~km~s$^{-1}$ and a radius of $0\fdg24$ that is compatible with the shell radius of SN~1006.

\subsection{Mass and density of the H{\sc i} clouds}\label{results:mass}
To derive the mass of the H{\sc i} clouds $M_\mathrm{HI}$ at $V_\mathrm{LSR}$: 4.0--12.0~km~s$^{-1}$, we used the following equation:
\begin{eqnarray}
M_\mathrm{HI} = m_{\mathrm{p}} \Omega D^2 \sum_{i} N_i(\mathrm{H}{\textsc{i}}),
\label{eq1}
\end{eqnarray}
where $m_\mathrm{p}$ is the mass of hydrogen, $\Omega$ is the solid angle for each data pixel, $D$ is the distance to the SNR, and $N$(H{\sc i}) is the atomic hydrogen column density. In general, $N$(H{\sc i}) can be derived as $1.823 \times W$(H{\sc i}), where $W$(H{\sc i}) is the H{\sc i} integrated intensity. Note that equation (\ref{eq1}) is valid for the optical depth of H{\sc i}$ \ll 1$. However, the latest observational and theoretical studies indicate that almost all H{\sc i} clouds are optically thick \citep[e.g.,][]{2014ApJ...796...59F,2015ApJ...798....6F,2018ApJ...860...33F,2019ApJ...884..130H,2020A&A...634A..83W,2021arXiv210910917S}. According to \cite{2015ApJ...798....6F}, the optical-depth-corrected H{\sc i} column density $N\mathrm{p}$'(H{\sc i}) is typically twice higher than $N$(H{\sc i}) calculated using equation (\ref{eq1}). Since the result was derived using the dust opacity map at 353 GHz \citep{2014A&A...571A..11P} toward the intermediate- and high {galactic} latitude clouds, this is applicable to SN~1006 at the intermediate latitude of $\sim$15$^{\circ}$. {Here,} we use a relation presented by \cite{2015ApJ...798....6F,2017ApJ...850...71F} that derives $N_\mathrm{p}$'(H{\sc i}) as a function of $W$(H{\sc i}). We then calculated the mass of the H{\sc i} clouds within the shell radius of $0\fdg24$ \citep[or $\sim$9 pc,][]{2010A&A...516A..62A} is $\sim$1000 $M_\odot$ and the averaged atomic hydrogen column density is $\sim$$4 \times 10^{20}$~cm$^{-2}$.

\section{Discussion}\label{discussion}
\subsection{Atomic Hydrogen Gas Associated with SN~1006}\label{discussion:associated}
\cite{2014ApJ...782L..33M} proposed that the southwest H{\sc i} cloud peaked at $\sim$8~km~s$^{-1}$ is interacting with the SNR, by comparing spatial distributions of the H{\sc i} cloud, the indentation of the X-ray shell, and the cutoff energy of synchrotron emission. Here, we suggest that the H{\sc i} clouds at $V_\mathrm{LSR} = 4.0$--12.0~km~s$^{-1}$ are most likely associated with the SNR from a kinematic point of view.

We first argue that the cavity-like distributions of H{\sc i} in the p--v and r--v diagrams provide us with a hint for the physical association with the atomic hydrogen gas at the velocity range of 4.0--12.0~km~s$^{-1}$. Because such cavity-like distributions in an SNR represent an expanding gas, and they are thought to be formed by shockwaves and/or strong winds from the progenitor system of the SNR \citep[e.g.,][]{1990ApJ...364..178K,1991ApJ...382..204K,1999ApJ...522..487H,1999ApJ...519..314H}. In the case of SN~1006, the expansion velocity is $\sim$4~km~s$^{-1}$ centered at the systemic velocity of $8 \pm 2$~km~s$^{-1}$. It is also noteworthy that the projected wind-shell gives the maximum extent near the systemic velocity, where we find a hollowed-out distribution of H{\sc i} as shown in Figure \ref{fig2}(c). Moreover, the maximum spatial extent of the expanding shell is found to be roughly the same size of the SNR shell as shown in Figure \ref{fig3}. This indicates that the forward shock has already reached the wind-shell, because the free expansion phase inside the shell is short enough owing to a much lower density \citep[e.g.,][]{1977ApJ...218..377W}. {In fact, \cite{2007ApJ...662..472B} have already predicted such a situation using the one-dimensional numerical simulation.} This can naturally explain the indentation of the X-ray shell toward the southwest H{\sc i} cloud suggested by \cite{2014ApJ...782L..33M}.

Next, we emphasize that the H{\sc i}-derived systemic velocity at $\sim$8~km~s$^{-1}$ coexists with the conventional source distance of 2.2~kpc. Although the systemic velocity at the distance of 2.2~kpc represents about --32~km~s$^{-1}$ by adopting the Galactic rotation curve model \citep{1993A&A...275...67B} with conventional Galactic parameters of $R_0 = 8.5$~kpc and $\Theta_0 = 220$~km s$^{-1}$~km~s$^{-1}$ \citep[IAU recommended values,][]{1986MNRAS.221.1023K}, the velocity difference about 40~km~s$^{-1}$ is not a problem since SN~1006 is placed almost 600~pc away from the Galactic plane. This implies that SN~1006 and its surrounding gas do not follow the Galactic rotation as also pointed out by \cite{2002A&A...387.1047D} and \cite{2014ApJ...782L..33M}.

The almost circular shape of SN~1006 without strong deformation is naturally expected by considering the column density of the shocked clouds \citep[e.g.,][]{2009ApJ...706L.106L,2017ApJS..230....2B}. In general, the shell morphology approaches a circular shape with decreasing the density of shock-associated clouds {\citep[e.g.,][]{2022MNRAS.512..265F}}. The young TeV gamma-ray SNR RX~J0852.0$-$4622 ($\sim$1700 yr) is a good example because the SNR shows almost circular shell. The total interstellar proton column density of shock-associated clouds is $\sim3 \times 10^{21}$~cm$^{-2}$ for RX~J0852.0$-$4622 \citep[][]{2017ApJ...850...71F,2018ApJ...866...76M}. By contrast, young ($\sim$1600 yr) TeV gamma-ray SNR RX~J1713.7$-$3946 shows a strongly deformed X-ray shell owing to shock-interactions with dense clouds of $\sim7 \times 10^{21}$~cm$^{-2}$ as averaged column density \citep[e.g.,][]{2003PASJ...55L..61F,2012ApJ...746...82F,2021ApJ...915...84F,2010ApJ...724...59S,2013ApJ...778...59S,2015ApJ...799..175S}. In the case of SN~1006, the column density of the shocked H{\sc i} cloud is $\sim4 \times 10^{20}$~cm$^{-2}$ (see Section \ref{results:mass}). Because the cloud density in SN~1006 is at least one order magnitude smaller than that in the three similar SNRs, the almost circular shape of SN~1006 is expected even if the shock-cloud interactions occurred. 

{Moreover, the previous proper-motion measurements may also be consistent with the H{\sc i} distributions at $V_\mathrm{LSR} = 4.0$--12.0~km~s$^{-1}$. According to \cite{2014ApJ...781...65W}, the highest velocity of $\sim$$7400\pm800$ km s$^{-1}$ was found in the southeast shell where no dense H{\sc i} clouds are located (see Figure \ref{fig1}b). On the other hand, the slower shock velocities of $\sim$5000 km s$^{-1}$ are seen in the northeast and southwest shells with rich H{\sc i} clouds (see also Figure 1b). By considering the forward shock interaction with the inner wall of the H{\sc i} shell, we can possibly find rapid deceleration of the shock wave toward the northeast and southwest shells of SN~1006 \citep[e.g.,][]{2021ApJ...906L...3T}.}

In conclusion, we claim that the H{\sc i} clouds at $V_\mathrm{LSR} = 4.0$--12.0~km~s$^{-1}$ are likely associated with SN~1006 in terms of their spatial distributions, kinetics, and physical properties.

\subsection{A Hint for a Single Degenerate Origin}\label{discussion:sd}
As described in Section \ref{results:mass}, the expanding H{\sc i} shell associated with SN~1006 has a mass of $\sim$1000 $M_\odot$. If the ambient medium with this large mass was uniformly distributed over the present volume of the remnant before being blown out, the initial ambient density is estimated to be $\sim$12~cm$^{-3}$ (here we assumed the shell radius of $\sim$9 pc \citep{2010A&A...516A..62A}). On the other hand, previous X-ray studies indicated the low pre-shock density of $\sim$0.02--0.4~cm$^{-3}$, based on the high velocity of the SNR forward shock \citep[e.g.,][]{2009ApJ...692L.105K,2014ApJ...781...65W} and low ionization state of the post-shock ISM and Fe ejecta \citep[][]{2007A&A...475..883A,2014ApJ...785L..27Y}. This discrepancy implies that the expanding H{\sc i} shell was first formed by the strong pre-explosion winds and subsequently the progenitor of SN~1006 exploded inside the low-density cavity. 

Because such wind activity prior to a Type Ia supernova explosion is {thought to be} associated with the SD scenario, we discuss whether the typical SD progenitor system can form the expanding H{\sc i} shell discovered in SN~1006. \cite{1999ApJ...522..487H,1999ApJ...519..314H} presented that the typical wind mass-loss rate is $\sim$$2 \times 10^{-6}$ $M_\odot$ yr$^{-1}$ (up to $\sim$$10^{-4}$ $M_\odot$ yr$^{-1}$, see also \citeauthor{2005ASPC..342..105N} \citeyear{2005ASPC..342..105N}) and the wind velocity is $\sim$2000~km~s$^{-1}$. If we adopt the dynamical time scale of expanding H{\sc i} shell as the wind duration period, we derive the momentum of accretion winds to be $\sim$8000 $M_\odot$~km~s$^{-1}$ or more. On the other hand, the momentum of expanding H{\sc i} shell is to be $\sim$4000 $M_\odot$~km~s$^{-1}$, by adopting the expansion velocity of $\sim$4~km~s$^{-1}$ and the H{\sc i} cloud mass of $\sim$1000 $M_\odot$. Therefore, the SD scenario can adequately explain the momentum of the observed expanding H{\sc i} shell.

{Finally, we discuss whether only the SD channel can produce the optically-thick winds through a phase of accreting material from a companion. According to \cite{2013A&ARv..21...59I}, the DD channel also undergoes several phases in their evolution that are not clear, in particular, the ``common envelope phase.'' The DD channel also experiences stages of accretion but maybe not stable enough or extended sufficiently in time compared to the SD channel. Since there are phases that we do not understand well in the DD channel, this uncertainty is a limitation of the present study to distinguish the SD and DD models. Another possibility is that a red supergiant with strong stellar winds happened to be in the line of sight. This possibility has been eliminated by the previous dedicated studies of a companion star search \citep[e.g.,][]{1980ApJ...241.1039S,2012Natur.489..533G,2012ApJ...759....7K,2018MNRAS.479..192K}. In any case, we would emphasize that the present H{\sc i} results and current knowledge also favor the SD scenario as the explosion mechanism of SN~1006, nevertheless, no surviving companion has been detected.}

\begin{figure*}[]
\begin{center}
\includegraphics[width=145mm]{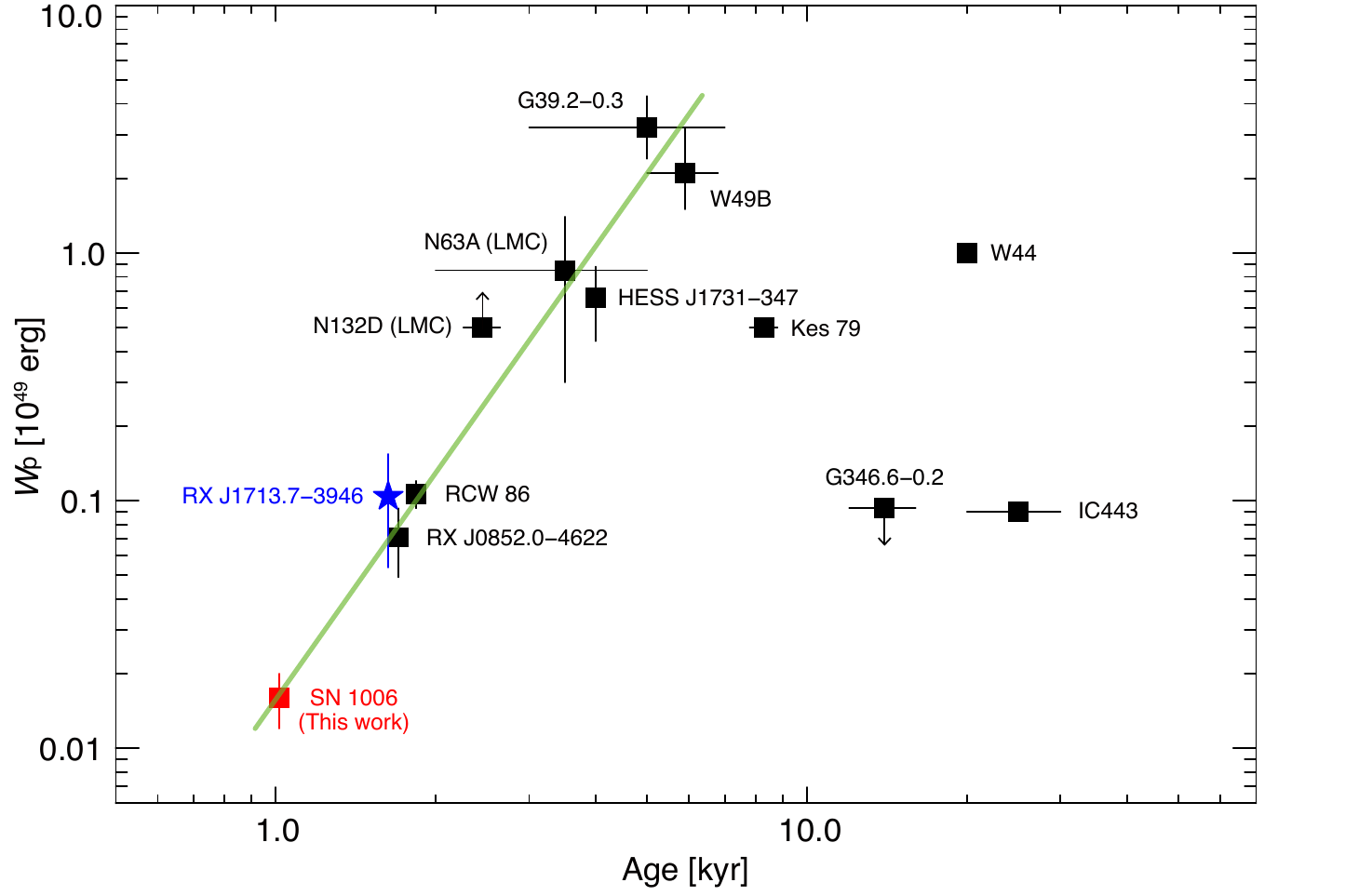}
\caption{Scatter plot between the age of SNRs and the total energy of cosmic-ray protons $W_\mathrm{p}$ \citep{2021ApJ...919..123S}. The green line indicates the linear regression of the double-logarithmic plot applying the least-squares fitting for data points with the ages of SNRs below 6 kyr. {The hadronic gamma-ray luminocity for each SNR was derived from the previous SED modeling alone except for RX~J1713.7$-3946$ (see the text).}}
\label{fig4}
\end{center}
\end{figure*}%

\begin{deluxetable*}{lcccccl}[]
\tablecaption{Comparison of Physical Properties in 13 Gamma-Ray SNRs}
\tablehead{
\multicolumn{1}{c}{Name}  & Distance & Diameter & Age & $n_\mathrm{p}$ & $W_\mathrm{p}$ & \multicolumn{1}{c}{References} \\
 & (kpc) & (pc) & (kyr) & (cm$^{-3}$) & ($10^{49}$~erg) & \\
 \multicolumn{1}{c}{(1)} & (2) & (3) & (4) & (5) & (6) & \multicolumn{1}{c}{(7)}}
\startdata
SN~1006 & \phantom{0}2.2$^\mathrm{a}$\ & 18 & 1.0 & \phantom{0}25 & \phantom{0iiiii}$0.016^{+0.004}_{-0.004}$ & This work \\
RX~J1713.7$-3946$ & 1.0 & 18 & 1.6 & 130 & {\phantom{00}$0.10^{+0.05}_{-0.05}$} & {\cite{2021ApJ...915...84F}} \\
RX~J0852.0$-$4622 & \phantom{0z}0.75$^\mathrm{b}$ & 24 & \phantom{z}1.7$^\mathrm{a}$ & 100 &  \phantom{00}$0.07^{+0.02}_{-0.02}$  & \cite{2017ApJ...850...71F} \\
RCW~86 & 2.5 & 30 & 1.8 & \phantom{0}75 & \phantom{00}$0.11^{+0.01}_{-0.01}$ & \cite{2019ApJ...876...37S} \\
HESS~J1731$-$347 & 5.7 & 44 & 4.0&  \phantom{0}60 & \phantom{00}$0.66^{+0.22}_{-0.22}$ & \cite{2014ApJ...788...94F} \\
G39.2$-$0.3 & 6.2 & 14 & \phantom{0}\phantom{0zw}$5.0^{+2.0}_{-2.0}$$^\mathrm{c}$ & 400 & $3.2^{+1.1}_{-0.8}$ & \cite{2020MNRAS.497.3581D} \\
W49B & 11.0\phantom{0} & 16 &  \phantom{0}\phantom{0zw}$6.0^{+1.0}_{-1.0}$$^\mathrm{d}$ & 650 & $2.1^{+1.1}_{-0.6}$  & \cite{2021ApJ...919..123S} \\
Kes~79 & 5.5 & 16 & \phantom{0zw}$8.3^{+0.5}_{-0.5}$ & 360 & 0.5\phantom{0zw} & \cite{2018ApJ...864..161K} \\
G346.6$-$0.2 & 11.1\phantom{0}  & 21 & \phantom{0iiiii}$14.0^{+2.0}_{-2.0}$ & 280 & $< 0.09$ &  \cite{2021arXiv210803392S}\\
W44 & \phantom{0}3.0$^\mathrm{e}$ & 27 & 20.0$^\mathrm{f}$& 200 & 1.0\phantom{0zw} &  \cite{2013ApJ...768..179Y} \\
IC443 & \phantom{j}1.5$^\mathrm{g}$ & 20 & \phantom{0iiiii}$25.0^{+5.0}_{-5.0}$$^\mathrm{h}$  & 680 & 0.09\phantom{ccz} & \cite{2022MNRASsubmitted} \\
\hline
LMC N132D & 50.0\phantom{0} & 25 & \phantom{00zw}$2.5^{+0.2}_{-0.2}$$^\mathrm{j}$ & $< 2000$& $> 0.5$ & \cite{2020ApJ...902...53S} \\
LMC N63A & 50.0\phantom{0} &  18 & \phantom{00zii}$3.5^{+1.5}_{-1.5}$$^\mathrm{j}$ & 190 & $0.9^{+0.5}_{-0.6}$ & \cite{2019ApJ...873...40S} \\
\enddata
\label{tab1}
\tablecomments{Col. (1): Name of SNRs. Col. (2): Distance to SNRs in units of kpc. Col. (3): Diameter of SNRs in units of pc. Col. (4): Age of SNRs in units of kyr. Col. (5): Averaged number density of total interstellar protons $n_\mathrm{p}$ in units of~cm$^{-3}$. Col. (6): Total energy of cosmic-ray protons $W_\mathrm{p}$ in units of 10$^{49}$ erg. Col. (7): References for CO/\ion{H}{1} derived $n_\mathrm{p}$ and $W_\mathrm{p}$ for each SNR. Other specific references are also shown as follows: $^\mathrm{a}$\cite{2003ApJ...585..324W}, $^\mathrm{b}$\cite{2008ApJ...678L..35K}, $^\mathrm{c}$\cite{2011ApJ...727...43S}, $^\mathrm{d}$\cite{2018AA...615A.150Z}, $^\mathrm{e}$\cite{1975AA....45..239C}, $^\mathrm{f}$\cite{1991ApJ...372L..99W}, $^\mathrm{g}$\cite{2003AA...408..545W}, $^\mathrm{h}$\cite{2008AJ....135..796L}; \cite{2001ApJ...554L.205O}, $^\mathrm{i}$\cite{2020ApJ...894...73L}, and $^\mathrm{j}$\cite{1998ApJ...505..732H}.}
\end{deluxetable*}

\subsection{Total Energy of Cosmic Ray Protons}\label{discussion:wp}
SNRs are thought to be promising acceleration sites for cosmic-ray protons, up to at least a few Peta electronvolts through the diffusive shock acceleration \citep[DSA, e.g.,][]{1978MNRAS.182..147B,1978ApJ...221L..29B}. By considering the injection rate of cosmic-rays and the total power of supernova explosions, the conventional value of the total energy of cosmic-rays $W_\mathrm{p}$ is to be $\sim$$10^{49}$--$10^{50}$ erg per a supernova explosion. Since $W_\mathrm{p}$ is proportional to the gamma-ray luminosity and the inverse of gas density, we can constrain the value of $W_\mathrm{p}$ for each SNR by observations. However, observational values of $W_\mathrm{p}$ still had large ambiguities because of the lack of unified quantification for shock-interacting molecular/atomic clouds through the CO/H{\sc i} radio line observations. 

Most recently, \cite{2021ApJ...919..123S,2021arXiv210803392S} summarized observational $W_\mathrm{p}$ values for 12 gamma-ray SNRs by adopting the number densities of shocked clouds using CO/H{\sc i} datasets. The authors found a tight relation between the SNR age and $W_\mathrm{p}$ for 12 gamma-ray SNRs: the young SNRs below $\sim$6 kyr show a positive correlation between them, while the older SNRs more than $\sim$8 kyr show a steady decrease of $W_\mathrm{p}$. The authors proposed that this trend can be explained as a combination of the age-limited acceleration \citep[e.g.,][]{2010A&A...513A..17O} and the energy-dependent diffusion of cosmic rays \citep[e.g.,][]{1996A&A...309..917A,2013ASSP...34..221G}. If the trend is real, SN~1006 shows a much lower value of $W_\mathrm{p}$ because of the low gamma-ray luminosity and gas density as well as its young age. In the present section, we derive the $W_\mathrm{p}$ value of SN~1006 and compare it with other gamma-ray SNRs.

According to the latest broad-band spectral modeling of SN~1006, almost gamma-ray emission is leptonic-dominated, which was produced from the inverse Compton scattering between accelerated cosmic-ray electrons and interstellar photons. On the other hand, hadronic gamma-rays, produced by interactions between cosmic-ray protons and interstellar protons, are thought to be contributed to gamma-rays from SN~1006. The total energy of accelerated cosmic-ray protons $W_\mathrm{p}$ is derived by \cite{2019PASJ...71...77X} as:
\begin{eqnarray}
W_\mathrm{p} = 1.5\mathrm{-}2.5 \times 10^{49} (n / 0.2 \;\mathrm{cm}^{-3})^{-1}\;\;\mathrm{erg}.
\label{eq2}
\end{eqnarray}
where $n$ is the number density of interstellar protons. In SN~1006, the averaged interstellar proton density is estimated to be $\sim$25~cm$^{-3}$ by adopting a shell radius of $0\fdg24$ or $\sim$9.2 pc and a shell thickness of $0\fdg05$ degree or $\sim$1.9 pc \citep[][]{2010A&A...516A..62A}. We then obtained $W_\mathrm{p} = 1.2$--$2.0 \times 10^{47}$ erg, which corresponds to $\sim$0.02\% of the typical released kinetic energy of a supernova explosion of $\sim$10$^{51}$ erg.

Figure \ref{fig4} shows the scatter plot between the SNR age and $W_\mathrm{p}$ for 13 gamma-ray SNRs that are listed in Table \ref{tab1}. {Note that the hadronic gamma-ray luminosity for deriving the $W_\mathrm{p}$ value in each SNR was calculated by the SED modeling alone except for RX~J1713.7$-$3946 \citep[see also][]{2021ApJ...915...84F}.} We find that SN~1006 lies on the regression line which was fitted using the data points with the ages of SNRs below 6 kyr, suggesting that the positive relation between the SNR age and $W_\mathrm{p}$ is applicable to gamma-ray SNRs with ages at least $\sim$1--6 kyr. If so, the $W_\mathrm{p}$ value of SN~1006 will increase up to several 10$^{49}$ erg in the next 5 kyr, even if the forward shock has already reached the wind-shell (see also Section \ref{discussion:associated}). Since it is unlikely that the decelerated forward shock would continue to accelerate cosmic rays for the next 5 kyr, some other mechanisms to increasing $W_\mathrm{p}$ are needed. 

One possible idea is that the cosmic-ray diffusion into the wind wall also plays an important role in understanding the values of $W_\mathrm{p}$ in the early evolutional stage of the SNRs. In this scenario, cosmic rays are mainly accelerated inside the low-density wind bubble via the DSA scheme. After the shock has reached the wind wall, cosmic rays diffuse into the wind-wall. The penetration depth of cosmic ray protons $l_\mathrm{pd}$ can be derived by \citep{2012ApJ...744...71I}:
\begin{eqnarray}
l_\mathrm{pd} =  0.1\; \eta^{0.5} (E / 10\;\mathrm{TeV})^{0.5} (B / 100\;\mathrm{\mu G})^{-0.5}\nonumber\\
(t/1000\;\mathrm{yr})^{0.5}\;\;\mathrm{pc}\;\;\;\;\;
\label{eq3}
\end{eqnarray}
where $\eta$ is gyro-factor ($> 1$), $E$ is the energy of cosmic rays, $B$ is the magnetic field, and $t$ is the age of the SNR. By adopting $\eta = 4$ for the inert-cloud region \citep{2020ApJ...900L...5T}, $E = 100$ TeV, and $B = 45$ $\mu$G \citep{2010A&A...516A..62A}, the penetration depth of cosmic-ray protons $l_\mathrm{pd}$ is to be $\sim$0.9 pc for $t = 1$ kyr and $\sim$2.3 pc for $t = 6$ kyr. Because the thickness of the wind shell is to be $\sim$1.9 pc, accelerated cosmic-ray protons will be fully interacting with the H{\sc i} clouds within the wind-shell in the next 5 kyr. In short, accelerated cosmic-rays below 100 TeV are trapped within the wind-cavity if the SNR age is young enough. 

{It should be also noted that the derived $W_\mathrm{p}$ values except for RX~J1713.7$-$3946 likely have additional uncertainties (by a factor of two or three) due to the difficulty in separation of the hadronic and leptonic gamma-rays by the SED modeling alone \citep[e.g.,][]{2012ApJ...744...71I}. According to \cite{2021ApJ...915...84F}, each gamma-ray component can be accurately distinguished by a comparison of gamma-ray, synchrotron X-ray, and total interstellar proton images. They found that hadronic gamma-ray contribution for RX~J1713.7$-$3946 is ($67 \pm 8$)\% of the total gamma-rays, and hence the accuracy of $W_\mathrm{p}$ (and the hadronic gamma-ray luminosity) in RX~J1713.7$-$3946 is significantly better than that in other SNRs derived using the SED modeling results. Moreover, all derived $W_\mathrm{p}$ values should be considered as an upper limit because we assume the uniform density distribution of the ISM protons within the shell. Nevertheless, we can find the global trend between the age and $W_\mathrm{p}$ by three orders of magnitude, implying that the trend itself is reliable.}

In any case, the young (age $< 6$ kyr) gamma-ray SNRs including SN~1006 show a good correlation between the SNR age and $W_\mathrm{p}$, possibly suggesting that the diffusion timescale is important in understanding in-situ values of $W_\mathrm{p}$. Further gamma-ray and H{\sc i} observations at the high-angular resolution using the Cherenkov Telescope Array \citep[CTA;][]{2011ExA....32..193A,2019scta.book.....C} and the Australian Square Kilometre Array Pathfinder \citep[ASKAP,][]{2007PASA...24..174J,2021PASA...38....9H} will allow us to reveal the diffusion mechanisms of cosmic rays in detail.

\section{Conclusions}\label{conclusions}
We summarize our conclusions as follows:
\begin{enumerate}
\item New H{\sc i} observations using the Australia Telescope Compact Array have revealed the spatial and kinematic distributions of H{\sc i} clouds associated with the Type Ia supernova remnant SN~1006. The H{\sc i} clouds at $V_\mathrm{LSR} = 4.0$--12.0~km~s$^{-1}$ show a good spatial correspondence with the X-ray shell, particularly in the southwest, northwest, and northeast. The total mass of H{\sc i} clouds is $\sim$1000 $M_\odot$ and the averaged atomic hydrogen column density is $\sim$$4 \times 10^{20}$~cm$^{-2}$ by assuming the optically thick H{\sc i}.
\item The H{\sc i} cavity-like distributions in the position--velocity and radius--velocity diagrams indicate the expanding shell, whose expansion velocity is $\sim$4~km~s$^{-1}$ with the systemic velocity of $8 \pm 2$~km~s$^{-1}$. By considering the pre- and post-shocked gas density and spatial extent of the expanding shell, the expanding H{\sc i} shell was likely formed by strong winds from the progenitor system, and then the forward shock of SN~1006 has already reached its wind wall. This scenario coexists with the conventional distance of 2.2~kpc because SN~1006 and its surroundings do not follow the Galactic rotation owing to their large distances from the Galactic plane.
\item We proposed a possible scenario that the progenitor system of SN~1006 consists of a white dwarf and a companion star, namely the single-degenerate system because the kinematics of the H{\sc i} expanding shell can be explained by accretion winds from the progenitors.
\item The total energy of accelerated cosmic-ray protons $W_\mathrm{p}$ is derived to be only $\sim$1.2--$2.0 \times 10^{47}$ erg by adopting the averaged interstellar proton density of $\sim$25~cm$^{-3}$. This small value is compatible with a positive correlation between the age and $W_\mathrm{p}$ of other gamma-ray supernova remnants with an age less than $\sim$6 kyr. Since the forward shock of SN~1006 has already reached the wind-shell and was decelerated, a time-dependent evolution of $W_\mathrm{p}$ is possibly relating the cosmic-ray diffusion into the H{\sc i} wind-shell. The cosmic-ray diffusion can increase the $W_\mathrm{p}$ value in SN~1006 up to several 10$^{49}$ erg in the next $\sim$5 kyr.
\end{enumerate}

\section*{Acknowledgements}
The authors acknowledge Tatsuya Fukuda and Satoshi Yoshiike for contributions on the data reduction and observations of H{\sc i}. {We are also grateful to the anonymous referee for useful comments that helped us improve the paper signiﬁcantly.} The Australia Telescope Compact Array (ATCA) and the Parkes radio telescope are parts of the Australia Telescope National Facility which is funded by the Australian Government for operation as a National Facility managed by CSIRO. We acknowledge the Gomeroi people as the traditional owners of the Observatory site. The scientific results reported in this article are based on data obtained from the Chandra Data Archive (Obs IDs: 3838, 4385--4394, 13738--13743, 14423, 14424, 14435). This research has made use of the software provided by the Chandra X-ray Center (CXC) in the application packages CIAO (v 4.12). This work was supported by JSPS KAKENHI Grant Numbers \href{https://kaken.nii.ac.jp/en/grant/KAKENHI-PUBLICLY-19H05075/}{JP19H05075} (H. Sano), \href{https://kaken.nii.ac.jp/en/grant/KAKENHI-PUBLICLY-20KK0309/}{JP20KK0309} (H. Sano), and \href{https://kaken.nii.ac.jp/en/grant/KAKENHI-PUBLICLY-21H01136/}{JP21H01136} (H. Sano).

\software{IDL Astronomy User's Library \citep{1993ASPC...52..246L}, MIRIAD \citep[][]{1995ASPC...77..433S}, CIAO \citep[v 4.12:][]{2006SPIE.6270E..1VF}, CALDB \citep[v 4.9.1][]{2007ChNew..14...33G}, KARMA \citep{1996ASPC..101...80G}.}

\facilities{Australia Telescope Compact Array (ATCA), Parkes, Chandra, High Energy Stereoscopic System (H.E.S.S.), Fermi.}


\begin{thebibliography}{99}
\bibitem[Acero et al.(2007)]{2007A&A...475..883A} Acero, F., Ballet, J., \& Decourchelle, A.\ 2007, \aap, 475, 883. doi:10.1051/0004-6361:20077742
\bibitem[Acero et al.(2010)]{2010A&A...516A..62A} Acero, F., Aharonian, F., Akhperjanian, A.~G., et al.\ 2010, \aap, 516, A62. doi:10.1051/0004-6361/200913916
\bibitem[Actis et al.(2011)]{2011ExA....32..193A} Actis, M., Agnetta, G., Aharonian, F., et al.\ 2011, Experimental Astronomy, 32, 193. doi:10.1007/s10686-011-9247-0
\bibitem[Aharonian \& Atoyan(1996)]{1996A&A...309..917A} Aharonian, F.~A. \& Atoyan, A.~M.\ 1996, \aap, 309, 917
\bibitem[Alsaberi et al.(2019)]{2019Ap&SS.364..204A} {Alsaberi, R.~Z.~E., Barnes, L.~A., Filipovi{\'c}, M.~D., et al.\ 2019, \apss, 364, 204. doi:10.1007/s10509-019-3696-8}
\bibitem[Badenes et al.(2007)]{2007ApJ...662..472B} {Badenes, C., Hughes, J.~P., Bravo, E., et al.\ 2007, \apj, 662, 472. doi:10.1086/518022}
\bibitem[Bamba et al.(2003)]{2003ApJ...589..827B} Bamba, A., Yamazaki, R., Ueno, M., et al.\ 2003, \apj, 589, 827. doi:10.1086/374687
\bibitem[Bamba et al.(2008)]{2008PASJ...60S.153B} Bamba, A., Fukazawa, Y., Hiraga, J.~S., et al.\ 2008, \pasj, 60, S153. doi:10.1093/pasj/60.sp1.S153
\bibitem[Bedin et al.(2014)]{2014MNRAS.439..354B} {Bedin, L.~R., Ruiz-Lapuente, P., Gonz{\'a}lez Hern{\'a}ndez, J.~I., et al.\ 2014, \mnras, 439, 354. doi:10.1093/mnras/stt2460}
\bibitem[Bell(1978)]{1978MNRAS.182..147B} Bell, A.~R.\ 1978, \mnras, 182, 147. doi:10.1093/mnras/182.2.147
\bibitem[Blandford \& Ostriker(1978)]{1978ApJ...221L..29B} Blandford, R.~D. \& Ostriker, J.~P.\ 1978, \apjl, 221, L29. doi:10.1086/182658
\bibitem[Bozzetto et al.(2017)]{2017ApJS..230....2B} {Bozzetto, L.~M., Filipovi{\'c}, M.~D., Vukoti{\'c}, B., et al.\ 2017, \apjs, 230, 2. doi:10.3847/1538-4365/aa653c}
\bibitem[Brand \& Blitz(1993)]{1993A&A...275...67B} Brand, J. \& Blitz, L.\ 1993, \aap, 275, 67
\bibitem[Cassam-Chena{\"\i} et al.(2008)]{2008ApJ...680.1180C} Cassam-Chena{\"\i}, G., Hughes, J.~P., Reynoso, E.~M., et al.\ 2008, \apj, 680, 1180. doi:10.1086/588015
\bibitem[Caswell et al.(1975)]{1975AA....45..239C} Caswell, J.~L., Murray, J.~D., Roger, R.~S., et al.\ 1975, \aap, 45, 239
\bibitem[Chen et al.(2017)]{2017A&A...604A..13C} Chen, X., Xiong, F., \& Yang, J.\ 2017, \aap, 604, A13. doi:10.1051/0004-6361/201630003
\bibitem[Cherenkov Telescope Array Consortium et al.(2019)]{2019scta.book.....C} Cherenkov Telescope Array Consortium, Acharya, B.~S., Agudo, I., et al.\ 2019, Science with the Cherenkov Telescope Array (Singapore: World Scientific). doi:10.1142/10986
\bibitem[Condon et al.(2017)]{2017ApJ...851..100C} Condon, B., Lemoine-Goumard, M., Acero, F., et al.\ 2017, \apj, 851, 100. doi:10.3847/1538-4357/aa9be8
\bibitem[de O{\~n}a Wilhelmi et al.(2020)]{2020MNRAS.497.3581D} de O{\~n}a Wilhelmi, E., Sushch, I., Brose, R., et al.\ 2020, \mnras, 497, 3581. doi:10.1093/mnras/staa2045
\bibitem[Di Stefano et al.(2011)]{2011ApJ...738L...1D} Di Stefano, R., Voss, R., \& Claeys, J.~S.~W.\ 2011, \apjl, 738, L1. doi:10.1088/2041-8205/738/1/L1
\bibitem[Dubner et al.(2002)]{2002A&A...387.1047D} Dubner, G.~M., Giacani, E.~B., Goss, W.~M., et al.\ 2002, \aap, 387, 1047. doi:10.1051/0004-6361:20020365
\bibitem[Filipovi{\'c} et al.(2022)]{2022MNRAS.512..265F} {Filipovi{\'c}, M.~D., Payne, J.~L., Alsaberi, R.~Z.~E., et al.\ 2022, \mnras, 512, 265. doi:10.1093/mnras/stac210}
\bibitem[Fruscione et al.(2006)]{2006SPIE.6270E..1VF} Fruscione, A., McDowell, J.~C., Allen, G.~E., et al.\ 2006, \procspie, 6270, 62701V. doi:10.1117/12.671760
\bibitem[Fukuda et al.(2014)]{2014ApJ...788...94F} Fukuda, T., Yoshiike, S., Sano, H., et al.\ 2014, \apj, 788, 94. doi:10.1088/0004-637X/788/1/94
\bibitem[Fukui et al.(2003)]{2003PASJ...55L..61F} {Fukui, Y., Moriguchi, Y., Tamura, K., et al.\ 2003, \pasj, 55, L61. doi:10.1093/pasj/55.5.L61}
\bibitem[Fukui et al.(2012)]{2012ApJ...746...82F} Fukui, Y., Sano, H., Sato, J., et al.\ 2012, \apj, 746, 82. doi:10.1088/0004-637X/746/1/82
\bibitem[Fukui et al.(2014)]{2014ApJ...796...59F} Fukui, Y., Okamoto, R., Kaji, R., et al.\ 2014, \apj, 796, 59. doi:10.1088/0004-637X/796/1/59
\bibitem[Fukui et al.(2015)]{2015ApJ...798....6F} Fukui, Y., Torii, K., Onishi, T., et al.\ 2015, \apj, 798, 6. doi:10.1088/0004-637X/798/1/6
\bibitem[Fukui et al.(2017)]{2017ApJ...850...71F} Fukui, Y., Sano, H., Sato, J., et al.\ 2017, \apj, 850, 71. doi:10.3847/1538-4357/aa9219
\bibitem[Fukui et al.(2018)]{2018ApJ...860...33F} Fukui, Y., Hayakawa, T., Inoue, T., et al.\ 2018, \apj, 860, 33. doi:10.3847/1538-4357/aac16c
\bibitem[Fukui et al.(2021)]{2021ApJ...915...84F} Fukui, Y., Sano, H., Yamane, Y., et al.\ 2021, \apj, 915, 84. doi:10.3847/1538-4357/abff4a
\bibitem[Fukushima et al.(2020)]{2020ApJ...897...62F} Fukushima, K., Yamaguchi, H., Slane, P.~O., et al.\ 2020, \apj, 897, 62. doi:10.3847/1538-4357/ab94a6
\bibitem[Gabici(2013)]{2013ASSP...34..221G} Gabici, S.\ 2013, Cosmic Rays in Star-Forming Environments (Berlin: Springer), 221. doi:10.1007/978-3-642-35410-6\_16
\bibitem[Gonz{\'a}lez Hern{\'a}ndez et al.(2009)]{2009ApJ...691....1G} {Gonz{\'a}lez Hern{\'a}ndez, J.~I., Ruiz-Lapuente, P., Filippenko, A.~V., et al.\ 2009, \apj, 691, 1. doi:10.1088/0004-637X/691/1/1}
\bibitem[Gonz{\'a}lez Hern{\'a}ndez et al.(2012)]{2012Natur.489..533G} Gonz{\'a}lez Hern{\'a}ndez, J.~I., Ruiz-Lapuente, P., Tabernero, H.~M., et al.\ 2012, \nat, 489, 533. doi:10.1038/nature11447
\bibitem[Gooch(1996)]{1996ASPC..101...80G} Gooch, R. 1996, in ASP Conf. Ser., 101, Karma: a Visualization Test-Bed, ed.
G. H. Jacoby \& J. Barnes (San Francisco, CA: ASP), 80
\bibitem[Graessle et al.(2007)]{2007ChNew..14...33G} Graessle, D.~E., Evans, I.~N., Glotfelty, K., et al.\ 2007, Chandra Newsletter, Vol. 14, p.33, 14
\bibitem[Hachisu et al.(1996)]{1996ApJ...470L..97H} Hachisu, I., Kato, M., \& Nomoto, K.\ 1996, \apjl, 470, L97. doi:10.1086/310303
\bibitem[Hachisu et al.(1999a)]{1999ApJ...522..487H} Hachisu, I., Kato, M., \& Nomoto, K.\ 1999a, \apj, 522, 487. doi:10.1086/307608
\bibitem[Hachisu et al.(1999b)]{1999ApJ...519..314H} Hachisu, I., Kato, M., Nomoto, K., et al.\ 1999b, \apj, 519, 314. doi:10.1086/307370
\bibitem[Hachisu \& Kato(2003a)]{2003ApJ...590..445H} Hachisu, I. \& Kato, M.\ 2003a, \apj, 590, 445. doi:10.1086/374968
\bibitem[Hachisu \& Kato(2003b)]{2003ApJ...598..527H} Hachisu, I. \& Kato, M.\ 2003b, \apj, 598, 527. doi:10.1086/378848
\bibitem[Hachisu et al.(2008)]{2008ApJ...679.1390H} Hachisu, I., Kato, M., \& Nomoto, K.\ 2008, \apj, 679, 1390. doi:10.1086/586700
\bibitem[Hayashi et al.(2019)]{2019ApJ...884..130H} Hayashi, K., Mizuno, T., Fukui, Y., et al.\ 2019, \apj, 884, 130. doi:10.3847/1538-4357/ab4351
\bibitem[Hachisu et al.(2012)]{2012ApJ...744...69H} Hachisu, I., Kato, M., Saio, H., et al.\ 2012, \apj, 744, 69. doi:10.1088/0004-637X/744/1/69
\bibitem[Hotan et al.(2021)]{2021PASA...38....9H} Hotan, A.~W., Bunton, J.~D., Chippendale, A.~P., et al.\ 2021, \pasa, 38, e009. doi:10.1017/pasa.2021.1
\bibitem[Hughes et al.(1998)]{1998ApJ...505..732H} Hughes, J.~P., Hayashi, I., \& Koyama, K.\ 1998, \apj, 505, 732. doi:10.1086/306202
\bibitem[Iben \& Tutukov(1984)]{1984ApJS...54..335I} Iben, I. \& Tutukov, A.~V.\ 1984, \apjs, 54, 335. doi:10.1086/190932
\bibitem[Inoue et al.(2012)]{2012ApJ...744...71I} Inoue, T., Yamazaki, R., Inutsuka, S.-. ichiro ., et al.\ 2012, \apj, 744, 71. doi:10.1088/0004-637X/744/1/71
\bibitem[Ivanova et al.(2013)]{2013A&ARv..21...59I} {Ivanova, N., Justham, S., Chen, X., et al.\ 2013, \aapr, 21, 59. doi:10.1007/s00159-013-0059-2}
\bibitem[Johnston et al.(2007)]{2007PASA...24..174J} Johnston, S., Bailes, M., Bartel, N., et al.\ 2007, \pasa, 24, 174. doi:10.1071/AS07033
\bibitem[Justham(2011)]{2011ApJ...730L..34J} Justham, S.\ 2011, \apjl, 730, L34. doi:10.1088/2041-8205/730/2/L34
\bibitem[Kalberla et al.(2010)]{2010A&A...521A..17K} Kalberla, P.~M.~W., McClure-Griffiths, N.~M., Pisano, D.~J., et al.\ 2010, \aap, 521, A17. doi:10.1051/0004-6361/200913979
\bibitem[Katsuda et al.(2008)]{2008ApJ...678L..35K} Katsuda, S., Tsunemi, H., \& Mori, K.\ 2008, \apjl, 678, L35. doi:10.1086/588499
\bibitem[Katsuda et al.(2009)]{2009ApJ...692L.105K} Katsuda, S., Petre, R., Long, K.~S., et al.\ 2009, \apjl, 692, L105. doi:10.1088/0004-637X/692/2/L105
\bibitem[Kerr \& Lynden-Bell(1986)]{1986MNRAS.221.1023K} Kerr, F.~J. \& Lynden-Bell, D.\ 1986, \mnras, 221, 1023. doi:10.1093/mnras/221.4.1023
\bibitem[Kerzendorf et al.(2012)]{2012ApJ...759....7K} Kerzendorf, W.~E., Schmidt, B.~P., Laird, J.~B., et al.\ 2012, \apj, 759, 7. doi:10.1088/0004-637X/759/1/7
\bibitem[Kerzendorf et al.(2013)]{2013ApJ...774...99K} {Kerzendorf, W.~E., Yong, D., Schmidt, B.~P., et al.\ 2013, \apj, 774, 99. doi:10.1088/0004-637X/774/2/99}
\bibitem[Kerzendorf et al.(2018a)]{2018MNRAS.479..192K} Kerzendorf, W.~E., Strampelli, G., Shen, K.~J., et al.\ 2018a, \mnras, 479, 192. doi:10.1093/mnras/sty1357
\bibitem[Kerzendorf et al.(2018b)]{2018MNRAS.479.5696K} {Kerzendorf, W.~E., Long, K.~S., Winkler, P.~F., et al.\ 2018b, \mnras, 479, 5696. doi:10.1093/mnras/sty1863}
\bibitem[Kirshner et al.(1987)]{1987ApJ...315L.135K} Kirshner, R., Winkler, P.~F., \& Chevalier, R.~A.\ 1987, \apjl, 315, L135. doi:10.1086/184875
\bibitem[Koyama et al.(1995)]{1995Natur.378..255K} Koyama, K., Petre, R., Gotthelf, E.~V., et al.\ 1995, \nat, 378, 255. doi:10.1038/378255a0
\bibitem[Koo et al.(1990)]{1990ApJ...364..178K} Koo, B.-C., Reach, W.~T., Heiles, C., et al.\ 1990, \apj, 364, 178. doi:10.1086/169400
\bibitem[Koo \& Heiles(1991)]{1991ApJ...382..204K} Koo, B.-C. \& Heiles, C.\ 1991, \apj, 382, 204. doi:10.1086/170709
\bibitem[Kuriki et al.(2018)]{2018ApJ...864..161K} Kuriki, M., Sano, H., Kuno, N., et al.\ 2018, \apj, 864, 161. doi:10.3847/1538-4357/aad7be
\bibitem[Landsman(1993)]{1993ASPC...52..246L} Landsman, W.~B.\ 1993, Astronomical Data Analysis Software and Systems II, 52, 246
\bibitem[Law et al.(2020)]{2020ApJ...894...73L} Law, C.~J., Milisavljevic, D., Patnaude, D.~J., et al.\ 2020, \apj, 894, 73. doi:10.3847/1538-4357/ab873a
\bibitem[Lee et al.(2008)]{2008AJ....135..796L} Lee, J.-J., Koo, B.-C., Yun, M.~S., et al.\ 2008, \aj, 135, 796. doi:10.1088/0004-6256/135/3/796
\bibitem[Li et al.(2015)]{2015MNRAS.453.3953L} Li, J.-T., Decourchelle, A., Miceli, M., et al.\ 2015, \mnras, 453, 3953. doi:10.1093/mnras/stv1882
\bibitem[Li et al.(2018)]{2018ApJ...864...85L} Li, J.-T., Ballet, J., Miceli, M., et al.\ 2018, \apj, 864, 85. doi:10.3847/1538-4357/aad598
\bibitem[Long et al.(2003)]{2003ApJ...586.1162L} Long, K.~S., Reynolds, S.~P., Raymond, J.~C., et al.\ 2003, \apj, 586, 1162. doi:10.1086/367832
\bibitem[Lopez et al.(2009)]{2009ApJ...706L.106L} Lopez, L.~A., Ramirez-Ruiz, E., Badenes, C., et al.\ 2009, \apjl, 706, L106. doi:10.1088/0004-637X/706/1/L106
\bibitem[Maeda \& Terada(2016)]{2016IJMPD..2530024M} Maeda, K. \& Terada, Y.\ 2016, International Journal of Modern Physics D, 25, 1630024. doi:10.1142/S021827181630024X
\bibitem[Maxted et al.(2018)]{2018ApJ...866...76M} Maxted, N.~I., Filipovi{\'c}, M.~D., Sano, H., et al.\ 2018, \apj, 866, 76. doi:10.3847/1538-4357/aae082
\bibitem[Maoz et al.(2014)]{2014ARA&A..52..107M} {Maoz, D., Mannucci, F., \& Nelemans, G.\ 2014, \araa, 52, 107. doi:10.1146/annurev-astro-082812-141031}
\bibitem[McClure-Griffiths et al.(2009)]{2009ApJS..181..398M} McClure-Griffiths, N.~M., Pisano, D.~J., Calabretta, M.~R., et al.\ 2009, \apjs, 181, 398. doi:10.1088/0067-0049/181/2/398
\bibitem[Miceli et al.(2012)]{2012A&A...546A..66M} Miceli, M., Bocchino, F., Decourchelle, A., et al.\ 2012, \aap, 546, A66. doi:10.1051/0004-6361/201219766
\bibitem[Miceli et al.(2014)]{2014ApJ...782L..33M} Miceli, M., Acero, F., Dubner, G., et al.\ 2014, \apjl, 782, L33. doi:10.1088/2041-8205/782/2/L33
\bibitem[Nomoto et al.(2005)]{2005ASPC..342..105N} Nomoto, K., Suzuki, T., Deng, J., et al.\ 2005, in ASP Conf. Ser. 342, 1604-2005: Supernovae as Cosmological Lighthouses, ed. M. Turatto et al. (San Francisco, CA: ASP), 105
\bibitem[Nomoto(1982)]{1982ApJ...257..780N} Nomoto, K.\ 1982, \apj, 257, 780. doi:10.1086/160031
\bibitem[Ohira et al.(2010)]{2010A&A...513A..17O} Ohira, Y., Murase, K., \& Yamazaki, R.\ 2010, \aap, 513, A17. doi:10.1051/0004-6361/200913495
\bibitem[Okamoto et al.(2017)]{2017ApJ...838..132O} Okamoto, R., Yamamoto, H., Tachihara, K., et al.\ 2017, \apj, 838, 132. doi:10.3847/1538-4357/aa6747
\bibitem[Olbert et al.(2001)]{2001ApJ...554L.205O} Olbert, C.~M., Clearfield, C.~R., Williams, N.~E., et al.\ 2001, \apjl, 554, L205. doi:10.1086/321708
\bibitem[Paczynski(1985)]{1985ASSL..113....1P} Paczynski, B.\ 1985, in Cataclysmic Variables and Low-Mass X-ray Binaries, ed. D. Q. Lamb \& J. Patterson (Dordrecht: Reidel), 1. doi:10.1007/978-94-009-5319-2\_1
\bibitem[Perlmutter et al.(1999)]{1999ApJ...517..565P} Perlmutter, S., Aldering, G., Goldhaber, G., et al.\ 1999, \apj, 517, 565. doi:10.1086/307221
\bibitem[Planck Collaboration et al.(2014)]{2014A&A...571A..11P} Planck Collaboration, Abergel, A., Ade, P.~A.~R., et al.\ 2014, \aap, 571, A11. doi:10.1051/0004-6361/201323195
\bibitem[Raymond et al.(2007)]{2007ApJ...659.1257R} Raymond, J.~C., Korreck, K.~E., Sedlacek, Q.~C., et al.\ 2007, \apj, 659, 1257. doi:10.1086/512483
\bibitem[Ruiz-Lapuente(2019)]{2019NewAR..8501523R} {Ruiz-Lapuente, P.\ 2019, \nar, 85, 101523. doi:10.1016/j.newar.2019.101523}
\bibitem[Ruiz-Lapuente et al.(2019)]{2019ApJ...870..135R} {Ruiz-Lapuente, P., Gonz{\'a}lez Hern{\'a}ndez, J.~I., Mor, R., et al.\ 2019, \apj, 870, 135. doi:10.3847/1538-4357/aaf1c1}
\bibitem[Sano et al.(2010)]{2010ApJ...724...59S} Sano, H., Sato, J., Horachi, H., et al.\ 2010, \apj, 724, 59. doi:10.1088/0004-637X/724/1/59
\bibitem[Sano et al.(2013)]{2013ApJ...778...59S} Sano, H., Tanaka, T., Torii, K., et al.\ 2013, \apj, 778, 59. doi:10.1088/0004-637X/778/1/59
\bibitem[Sano et al.(2015)]{2015ApJ...799..175S} Sano, H., Fukuda, T., Yoshiike, S., et al.\ 2015, \apj, 799, 175. doi:10.1088/0004-637X/799/2/175
\bibitem[Sano et al.(2017)]{2017JHEAp..15....1S} Sano, H., Reynoso, E.~M., Mitsuishi, I., et al.\ 2017, Journal of High Energy Astrophysics, 15, 1. doi:10.1016/j.jheap.2017.04.002
\bibitem[Sano et al.(2018)]{2018ApJ...867....7S} Sano, H., Yamane, Y., Tokuda, K., et al.\ 2018, \apj, 867, 7. doi:10.3847/1538-4357/aae07c
\bibitem[Sano et al.(2019a)]{2019ApJ...876...37S} Sano, H., Rowell, G., Reynoso, E.~M., et al.\ 2019a, \apj, 876, 37. doi:10.3847/1538-4357/ab108f
\bibitem[Sano et al.(2019b)]{2019ApJ...873...40S} Sano, H., Matsumura, H., Nagaya, T., et al.\ 2019b, \apj, 873, 40. doi:10.3847/1538-4357/ab02fd
\bibitem[Sano et al.(2020a)]{2020ApJ...902...53S} Sano, H., Plucinsky, P.~P., Bamba, A., et al.\ 2020a, \apj, 902, 53. doi:10.3847/1538-4357/abb469
\bibitem[Sano et al.(2021a)]{2021ApJ...919..123S} Sano, H., Yoshiike, S., Yamane, Y., et al.\ 2021a, \apj, 919, 123. doi:10.3847/1538-4357/ac0dba
\bibitem[Sano et al.(2021b)]{2021arXiv210803392S} Sano, H., Suzuki, H., Nobukawa, K.~K., et al.\ 2021b, arXiv:2108.03392
\bibitem[Sault et al.(1995)]{1995ASPC...77..433S} Sault, R.~J., Teuben, P.~J., \& Wright, M.~C.~H.\ 1995, Astronomical Data Analysis Software and Systems IV, 77, 433
\bibitem[Schaefer(1996)]{1996ApJ...459..438S} Schaefer, B.~E.\ 1996, \apj, 459, 438. doi:10.1086/176906
\bibitem[Schweizer \& Middleditch(1980)]{1980ApJ...241.1039S} Schweizer, F. \& Middleditch, J.\ 1980, \apj, 241, 1039. doi:10.1086/158417
\bibitem[Seifried et al.(2021)]{2021arXiv210910917S} Seifried, D., Beuther, H., Walch, S., et al.\ 2021, arXiv:2109.10917
\bibitem[Stephenson \& Green(2002)]{2002ISAA....5.....S} Stephenson, F.~R. \& Green, D.~A.\ 2002, in Historical Supernovae and their Remnants, ed. F. R. Stephenson \& D. A. Green (Oxford: Clarendon), 5
\bibitem[Su et al.(2011)]{2011ApJ...727...43S} Su, Y., Chen, Y., Yang, J., et al.\ 2011, \apj, 727, 43. doi:10.1088/0004-637X/727/1/43
\bibitem[Tanaka et al.(2020)]{2020ApJ...900L...5T} Tanaka, T., Uchida, H., Sano, H., et al.\ 2020, \apjl, 900, L5. doi:10.3847/2041-8213/abaef0
\bibitem[Tanaka et al.(2021)]{2021ApJ...906L...3T} Tanaka, T., Okuno, T., Uchida, H., et al.\ 2021, \apjl, 906, L3. doi:10.3847/2041-8213/abd6cf
\bibitem[Uchida et al.(2013)]{2013ApJ...771...56U} Uchida, H., Yamaguchi, H., \& Koyama, K.\ 2013, \apj, 771, 56. doi:10.1088/0004-637X/771/1/56
\bibitem[Wang et al.(2020)]{2020A&A...634A..83W} Wang, Y., Beuther, H., Rugel, M.~R., et al.\ 2020, \aap, 634, A83. doi:10.1051/0004-6361/201937095
\bibitem[Weaver et al.(1977)]{1977ApJ...218..377W} Weaver, R., McCray, R., Castor, J., et al.\ 1977, \apj, 218, 377. doi:10.1086/155692
\bibitem[Webbink(1984)]{1984ApJ...277..355W} Webbink, R.~F.\ 1984, \apj, 277, 355. doi:10.1086/161701
\bibitem[Welsh \& Sallmen(2003)]{2003AA...408..545W} Welsh, B.~Y. \& Sallmen, S.\ 2003, \aap, 408, 545. doi:10.1051/0004-6361:20030908
\bibitem[Whelan \& Iben(1973)]{1973ApJ...186.1007W} Whelan, J. \& Iben, I.\ 1973, \apj, 186, 1007. doi:10.1086/152565
\bibitem[Winkler et al.(2003)]{2003ApJ...585..324W} Winkler, P.~F., Gupta, G., \& Long, K.~S.\ 2003, \apj, 585, 324. doi:10.1086/345985
\bibitem[Winkler et al.(2013)]{2013ApJ...764..156W} Winkler, P.~F., Williams, B.~J., Blair, W.~P., et al.\ 2013, \apj, 764, 156. doi:10.1088/0004-637X/764/2/156
\bibitem[Winkler et al.(2014)]{2014ApJ...781...65W} Winkler, P.~F., Williams, B.~J., Reynolds, S.~P., et al.\ 2014, \apj, 781, 65. doi:10.1088/0004-637X/781/2/65
\bibitem[Wolszczan et al.(1991)]{1991ApJ...372L..99W} Wolszczan, A., Cordes, J.~M., \& Dewey, R.~J.\ 1991, \apjl, 372, L99. doi:10.1086/186033
\bibitem[Woods et al.(2017)]{2017NatAs...1..800W} {Woods, T.~E., Ghavamian, P., Badenes, C., et al.\ 2017, Nature Astronomy, 1, 800. doi:10.1038/s41550-017-0263-5}
\bibitem[Xing et al.(2016)]{2016ApJ...823...44X} Xing, Y., Wang, Z., Zhang, X., et al.\ 2016, \apj, 823, 44. doi:10.3847/0004-637X/823/1/44
\bibitem[Xing et al.(2019)]{2019PASJ...71...77X} Xing, Y., Wang, Z., Zhang, X., et al.\ 2019, \pasj, 71, 77. doi:10.1093/pasj/psz056
\bibitem[Xue \& Schaefer(2015)]{2015ApJ...809..183X} {Xue, Z. \& Schaefer, B.~E.\ 2015, \apj, 809, 183. doi:10.1088/0004-637X/809/2/183}
\bibitem[Xue et al.(2021)]{2021MNRAS.501..664X} Xue, L., Jiao, C.-L., \& Li, Y.\ 2021, \mnras, 501, 664. doi:10.1093/mnras/staa3696
\bibitem[Yamaguchi et al.(2008)]{2008PASJ...60S.141Y} Yamaguchi, H., Koyama, K., Katsuda, S., et al.\ 2008, \pasj, 60, S141. doi:10.1093/pasj/60.sp1.S141
\bibitem[Yamaguchi et al.(2014)]{2014ApJ...785L..27Y} Yamaguchi, H., Badenes, C., Petre, R., et al.\ 2014, \apjl, 785, L27. doi:10.1088/2041-8205/785/2/L27
\bibitem[Yoshiike et al.(2013)]{2013ApJ...768..179Y} Yoshiike, S., Fukuda, T., Sano, H., et al.\ 2013, \apj, 768, 179. doi:10.1088/0004-637X/768/2/179
\bibitem[Yoshiike et al.(2022)]{2022MNRASsubmitted} Yoshiike, S., Sano, H., Fukuda, T.,  et al.\ 2022, to be submitted
\bibitem[Zhou et al.(2016)]{2016ApJ...826...34Z} Zhou, P., Chen, Y., Zhang, Z.-Y., et al.\ 2016, \apj, 826, 34. doi:10.3847/0004-637X/826/1/34
\bibitem[Zhou \& Vink(2018)]{2018AA...615A.150Z} Zhou, P. \& Vink, J.\ 2018, \aap, 615, A150. doi:10.1051/0004-6361/201731583
\end{thebibliography}
\end{document}